\documentstyle[prl,twocolumn,aps,graphics]{revtex}
\begin{document}
\draft
\twocolumn[\hsize\textwidth\columnwidth\hsize\csname @twocolumnfalse\endcsname
\title{Fractal Spin Glass Properties of Low Energy Configurations\\
 in the Frenkel-Kontorova chain}

\author{O.V. Zhirov\( ^{a}\), G. Casati\( ^{b } \), and D.L. Shepelyansky\( ^{c} \)}

\address {\( ^{a} \) Budker Institute of Nuclear Physics, 630090 Novosibirsk, Russia}

\address{\( ^{b} \) International Center for the Study of Dynamical Systems, 22100
Como, and \\
Istituto Nazionale di Fisica della Materia and INFN, Unita'di Milano,
Italy}

\address{\( ^{c} \) Laboratoire de Physique Quantique, UMR C5926 du CNRS, Universit\'e
Paul Sabatier, 31062 Toulouse, France}


\date{July 6, 2001}

\maketitle

\begin{abstract}
We study numerically and analytically the classical one-dimensional Frenkel-Kontorova chain 
in the regime of pinned phase characterized by  phonon gap. Our results show the existence of exponentially 
many static equilibrium configurations which are exponentially close to the energy of the ground 
state. The energies of these configurations form a fractal quasi-degenerate band structure which 
is described on the basis of elementary excitations. Contrary to the ground state, 
the configurations inside these bands are disordered.
\end{abstract}
\vskip1pc]

\narrowtext

The Frenkel-Kontorova  (FK) model \cite{FK38} 
describes a one-dimensional chain of atoms/particles with harmonic couplings 
placed in a periodic potential. This model was introduced more than sixty years
ago with the aim to study commensurate-incommensurate 
phase transitions \cite{FK38,Pokr84}. 
However, it was also successfully applied for the description  of 
crystal dislocations \cite{Naba67},
epitaxial monolayers on the crystal surface \cite{Ying71}, ionic conductors
and glassy materials \cite{Piet81,Au78,Au83a} and, more recently, to charge-density
waves \cite{Flor96} and dry friction \cite{Brau97,Cons00}. In addition, 
the FK model has also found its implementation in the investigation of the Josephson
junction chain \cite{Wata96}. 
The physical properties  of the FK model are very rich. Moreover, different types
 of interaction between atoms can be effectively reduced to the case of FK model
 and, due to that, this model continues to attract the active interest of 
different research groups.  

The ground state of the classical FK model is defined as the static, equilibrium
configuration of the chain, which corresponds to the \textit{absolute} minimum of
the chain potential energy. More than twenty years ago 
Aubry discovered \cite{Au78,Au83b,Au83c}
that the \textit{ground state} is unique and  is characterized by a special regular
order of atoms in the chain. In fact, the positions of atoms in the chain are described
by an area-preserving map which is well known in the field of dynamical
chaos and which is called the Chirikov standard map  \cite{Chir79}.
The density of particles in the FK model determines the rotation number
of the invariant curves of the map while the amplitude of the periodic potential
gives the value of the dimensionless parameter $K$. For  $K<K_{c}$ the KAM
curves are smooth and the spectrum of long wave phonon excitations in the chain
is characterized by a linear dispersion law starting from zero frequency. 
On the contrary for $K>K_{c}$ the KAM curves are destroyed and 
replaced by an invariant Cantor set which is called cantorus. In this regime 
the phonon
spectrum has a gap so that the phonon excitations are suppressed at
low temperature. The effects of the cantorus on the dynamical
properties of the map were discussed in \cite{Percival,Mackay}.
Later \cite{Vall86}, on the example of Ising spin model to which the 
FK model can be \textit{approximately} reduced \cite{Be80},
it has been shown that the ground state has some well defined hierarchical structure.
The main features of this structure are determined by the number properties of 
the dimensionless particle density which is given by the ratio
of the mean interparticle distance to the period of the external field.

In more recent studies \cite{Burk96,Keto97,Tong99,Hu00} the attention was 
mainly concentrated on
phonon modes in incommensurate one-dimensional chains. 
Indeed, the phonon modes contribute
to the specific heat of the system and hence they are responsible for the heat conduction
along the chain \cite{Tong99,Hu00}. 
The propagation and localization of phonon modes \cite{Burk96,Keto97} have been studied 
for  small vibrations of particles around their equilibrium positions 
in the \textit{ground state}.

However we would like to stress that for $K>K_c$, besides the ground state there
 exist other \textit{excited}  
equilibrium configurations, corresponding to local minima of the potential,
with energies very close to the ground state. To our knowledge only few studies 
were dedicated to excited equilibrium 
configurations, see for example \cite{Be80,Vall86,Au90}.
In this paper we study the properties of these low energy equilibrium configurations
and determine the structure of their energy spectrum and its dependence on
the strength of the periodic potential and on the chain length. The obtained results 
show that these configurations are exponentially close in energy to the ground state
and the number of configurations grows exponentially with the length of the chain.
We also show that these configurations have interesting fractal properties which we
will describe in detail. The transition between different configurations can be
understood on the basis of elementary excitations which we call
 \char`\"{}bricks\char`\"{}. The numerical and analytical study of these
elementary excitations allows to understand and describe the fractal structure
of energy bands corresponding to equilibrium configurations.
Since the excited equilibrium configurations  are exponentially close to the
ground state they will strongly contribute to the physical
system properties at finite temperature. Contrary to the ground state in which atoms form a regular structure,
in the excited configurations this order is partially destroyed and some chaotic feature
appears. In some sense the existence of an exponential number of configurations
 exponentially close in energy, reminds the situation in classical spin glasses
\cite{Parisi}. However, contrary to the usual spin glass models, the FK model
is described by a simple Hamiltonian without any disorder. Therefore the 
appearance of exponentially quasi-degenerate configurations in the FK system
can be viewed as a dynamical spin glass model. Thus the rich variety of properties 
of the FK model can find application in different areas of physics.

\section{The model.}

Let us consider a chain of particles with pairwise elastic interactions between
 nearest neighbors: \( V(x_{i},x_{i-1})=v(x_{i}-x_{i-1})=(x_{i}-x_{i-1})^{2}/2 \)
. This
chain is placed in a periodic external field: \( W(x_i)=-K\cos {(x_i)} \), where
(without any loss of generality) the period is taken equal to \( 2\pi  \). 
Therefore the Hamiltonian of the FK model reads:

\begin{equation}
\label{Ham}
H= \sum_i[{P_i^2 \over 2} + {(x_i -x_{i-1})^2 \over 2}- K \cos(x_i)].
\end{equation}

Here we have taken the mass of the particles and the elastic constant
equal to one. At the equilibrium the momenta $P_i =0$ and in addition

\begin{equation}
\label{equilib}
{\partial H \over \partial x_i} = -x_{i+1}+ 2 x_i - x_{i-1}+K\sin(x_i)=0.
\end{equation}

After the introduction of new variables $p_{i+1} = x_{i+1}-x_i$, this equation
can be written in the form of an area-preserving map:

\begin{equation}
\label{stmap}
p_{i+1}= p_i +K\sin(x_i) \;\; , \;\; x_{i+1}= x_i + p_{i+1} \; ,
\end{equation}

which is known as the Chirikov standard map \cite{Chir79}.

We concentrate our investigation on the case of golden mean dimensionless particle 
density $\nu=(\sqrt{5}-1)/2$. This irrational value can be approximated by rational 
approximants which form the Fibonacci sequence $s_n$ with number of particles $s$
and chain length $L=2 \pi r$ . In this way the rational approximants are
$\nu_n = r_n/s_n = s_{n-1}/s_n$, with $s_n = 1,2,3,5,8,13...$ and the average
distance between particles is $ a = 2 \pi \nu_n$.
For the map (\ref{stmap}) the parameter $\nu$ determines the rotation number
of the invariant Kolmogorov-Arnold-Moser (KAM) curve. At the golden mean
value of $\nu$ the KAM curve is analytical and smooth for $K<K_c = 0.971635..$
\cite{Gree79}. For $K>K_c$ the curve is destroyed and the transition by the
breaking of analyticity takes place\cite{Au78}. 
As a result the invariant curve is replaced by a cantorus which forms an invariant 
fractal set in the phase space of the map. For the FK model the cantorus
corresponds to the ground state with minimal energy as it was shown by Aubry
\cite{Au78,Au83a,Au83b,Au83c}.

In this paper we restrict ourselves to the case with \( K>K_{c} \), 
 when each particle is locked by potential barriers of the external
periodical field and the whole chain is pinned. Stable configurations of 
the chain correspond to minima of
the potential energy:

\begin{equation}
\label{Uchain}
U(\{x\})=\sum_i[{(x_i -x_{i-1})^2 \over 2}- K \cos(x_i)].
\end{equation}
 
The static ground state corresponds to \textit{absolute} minimum given by Aubry's
solution. However, as we will see in the next section, there are other local
minima of the chain potential $U(\{x\})$ which give equilibrium static configurations
 with energy being very close to the ground state. The number of such configurational 
states $N_{cs}$ grows exponentially with the chain length $s$.

\section{Energy spectrum of equilibrium configurations.\label{EnSpectra}}

\begin{figure}
{\par\centering \resizebox*{0.8\columnwidth}{!}{\rotatebox{90}{
\includegraphics{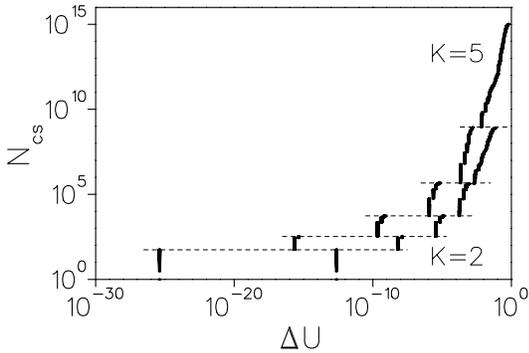}}} \par}
\vglue 0.2cm
\caption{\label{Idos}Integrated number of equilibrium configuration states
$N_{cs}$ as a function of the energy difference $\Delta U$ between the energy 
of configuration $U$ and the ground state energy $U_G$, counted per one 
particle. Here the number
of particles is \protect\( s=89\protect \) and the number of wells 
 $r=55$. The are shown for $K=5$ (6 upper segments) and for $K=2$ 
(5 lower segments). Horizontal dashed lines show the border between energy 
bands. All equilibrium configurations are counted.}
\end{figure}

In Fig.\ref{Idos} we present a typical result for the integrated number $N_{cs}$ 
of excited  equilibrium configurations  versus their energy
difference, per particle, from the ground state$\Delta U = (U -U_G)/s$ where $U_G$
is the energy of the ground state.   Here we have number of particles $s=89$, 
number of wells $r=55$ and two values of $K$,  $K=5$ and $K=2$.
This figure shows that the energy of configurations form a sequence of narrow energy bands the width of which is much smaller than the distance between bands, at least in the very vicinity of the ground state. At higher energies the band width starts to grow and eventually nearest bands almost merge into each other. It is interesting to note that the number of states in each band is practically independent of $K$ as it is shown by dashed lines in Fig.\ref{Idos}: with the increase of $K$ each band is shifted to smaller values of $\Delta U$ (in logarithmic scale) but the number of states in each band is not changed. 

It should be stressed, that even at a moderate value of \( K=2 \) the energy spacing between the ground state and the first excited configuration band is of the order of \( 10^{-13} \).
If one assumes that in Eq. (\ref{Uchain}) a  unit of energy is \( \sim 1 \)eV, then this band
is already excited at temperature  \(T \sim 10^{-9} \) \( ^{\circ } \)K! Hence
one may conclude, that the pure ground state is practically inaccessible even for
a chain with less than one hundred atoms.

The total
number of different local minima $N_{cs}$ is enormous and it grows very rapidly with  \( K \). Therefore to numerically find all these configurations one needs to use special methods. Our approach to this problem is the following.
First we find the ground state by the gradient method developed by Aubry \cite{Au83a,Pe83,Sche86}. Then we find the excited equilibrium configurations
with the help of the Metropolis algorithm \cite{Creu81}. In this method the system is considered at some properly chosen temperature $T$. At given $T$ we can probe the configurations with $\Delta U \leq T$ while the probability to find configurations with higher energy is exponentially suppressed.
Our implementation of the Metropolis algorithm looks as follows. We start from a certain configuration \( \{x\}_{j} \), which corresponds to some local minimum \( U_{j} \)
of the chain potential energy \( U(\{x\}_{j}) \). Then we take randomly one of the
chain particles and try to move it into one of the neighbour wells. Next, with
the new distribution of particles among the \textit{wells,} we search for a new
local minimum \( \widetilde{U} \). A new configuration with \( U_{j+1}=\widetilde{U} \)
is accepted if \( \exp (-(\widetilde{U}-U_{j})/T)\geq \xi  \), where \( \xi  \)
is a random number homogeneously distributed in the interval \( [0,1] \),
otherwise we try a new attempt. Notice that our Metropolis procedure uses
 particles \textit{jumps} from well to well rather than (small) variations of
their coordinates. In this way we solve the problem of the Peierls-Nabarro barriers \cite{Au78} and obtain a method with good performance. Physically the Peierls-Nabarro barriers are not important since we are interested in static configurations and not in the transition rate between different states.

In general the space of low-energy configurations can be viewed as a set of disconnected islands. Therefore there is a danger that, starting near one island, 
we can  remain in its vicinity forever. To avoid this we  periodically  heat/freeze  the system. During this process we perform the above described iterations with chosen temperature $T$. In this way the system can move from one island to  another and visit different equilibrium configurations.

Since the number of equilibrium configurations is exponentially large (see Fig .\ref{Idos}), it is not possible to visit and count exactly all of them. However their number can be counted approximately with sufficiently good accuracy in the following way.  In the lowest excited band the number of equilibrium configurations is not so large and it can be computed exactly. In order to determine the number of states in the next band we start from a representative sample of configurations, which is in fact a small part of their total number in one band. Then with the help of Metropolis algorithm described above we determine the ratio between the number of configurations inside the first and second band. To do this we choose the temperature value $T$ in such a way that  $T \sim 10 \Delta U_2 > \Delta U_1 $ where $\Delta U_{1,2}$ are the excitation energies for the first and second band counted from the ground state. From the computed ratio we determine with sufficiently good accuracy the total number of configurations in the second band. By iterating this process we determine the total number of states in all bands. Moreover, by gradually changing the temperature $T$, this procedure can be easily adapted to higher excitation energies when the bands begin to merge. This allows to compute the total number of equilibrium configurations in the system which, for $K=5$ is of the order of $10^{15}$.

\begin{figure}
{\par\centering \resizebox*{0.8\columnwidth}{!}{\rotatebox{90}{
\includegraphics{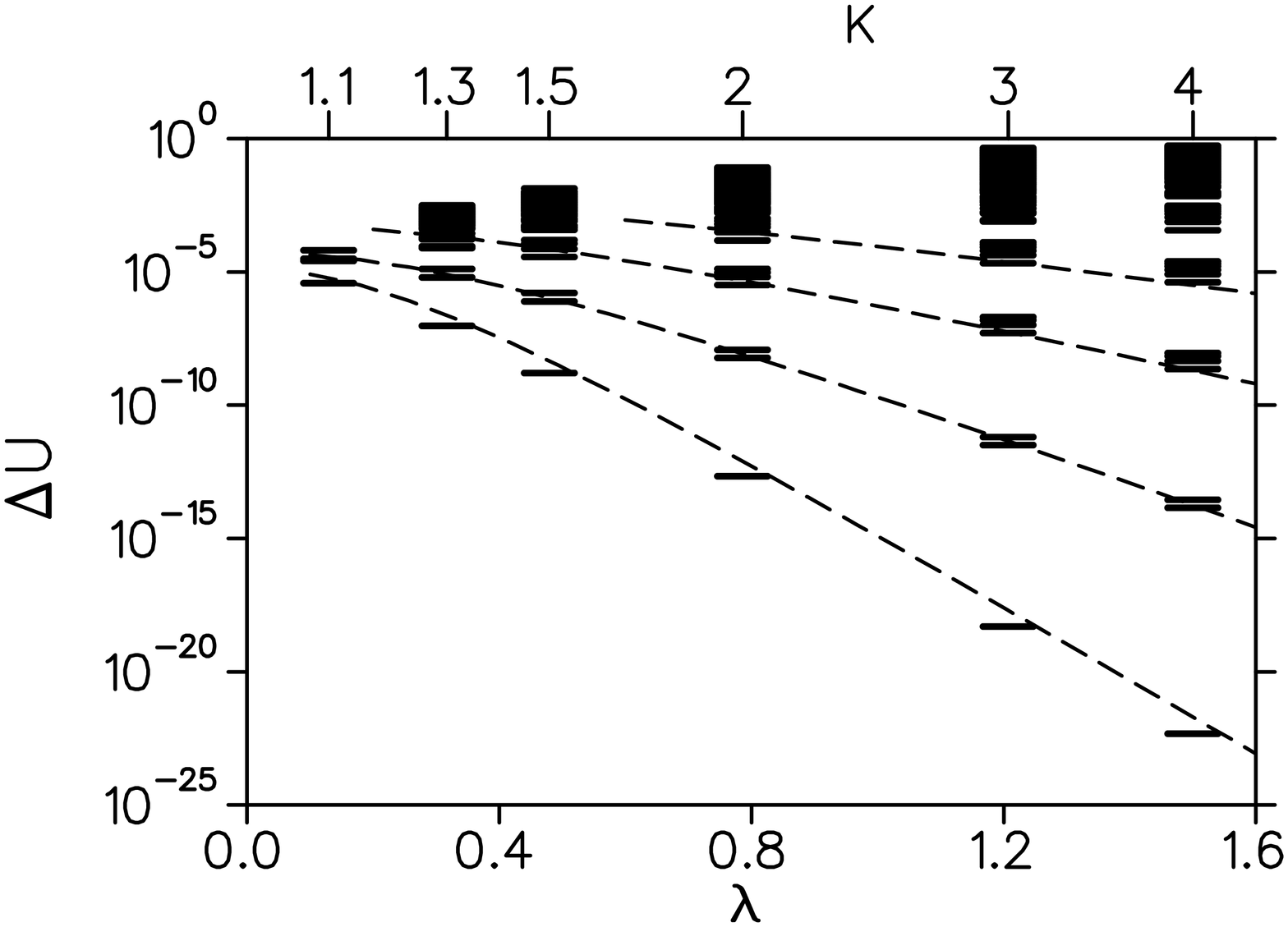}}} \par}
\vglue 0.2cm
\caption{\label{kdep}Band energy spectrum of equilibrium configurations versus the chaos parameter $K$ (upper scale) and the phonon gap $\lambda$ (lower scale). The bands are marked by filled areas corresponding to a given $K$ value. 
The chain is the same, as in Fig.\ref{Idos}: \protect\( r/s=55/89.\protect \). The dashed curves are given by the semi-empirical expression (\ref{bfit}).}
\end{figure}

The energy band spectrum for different values of $K$ is shown in Fig.\ref{kdep}. It clearly shows that the number of bands becomes larger for larger $K$ and, in addition, the lowest bands approach exponentially the ground state. The existence of such bands exponentially close to the ground state is related to the specific properties of the FK chain in the pinned phase ($K>K_c$). This phase is characterized by a phonon gap $\lambda$ \cite{Au78,Pe83} due to which any static displacement perturbation \( \delta x_{i_{0}} \) of particle $i_0$(corresponding to a zero-frequency {}``phonon{}'') decays exponentially along the chain: \( \delta x_{i}\propto \exp (-\lambda \cdot |i-i_{0}|) \).
In fact $\lambda$ is the Lyapunov exponent of the map (\ref{stmap}) computed on the cantorus. This exponential decay of perturbations is responsible for the appearance of exponentially narrow bands exponentially close to the ground state. 

In order to describe the band positions as a function of system parameters it is convenient to label the bands by the index $k$ in order of increasing energy. Then the energies of the four lowest bands are well described by a simple empirical formula, see Fig. \ref{kdep}:
 \begin{equation}
\label{bfit}
\left\langle \Delta U_k\right\rangle \simeq C\exp (-\alpha \nu ^{k}s\sqrt{\beta k+\lambda ^{2}}),
\end{equation}

where \( \left\langle \Delta U_k\right\rangle _{k} \) is the average energy of \( k \)-th
(excited) band, \( s \) is the number of particles in the chain, and the numerical values of parameters are \( C\approx 1 \),
\( \alpha \approx 0.59 \), \( \beta \approx 0.12 \). It is rather interesting to note 
that this simple formula describes quite well even the region with small values of \( \lambda \leq 0.8 \)
(\( K\leq 2 \)). At larger \( K \) (and longer chains) this formula can be
replaced by its even more simple limiting expression 

\begin{equation}
\label{bfitL}
\left\langle \Delta U_k\right\rangle\simeq C\exp (-\alpha \nu ^{k}s\lambda ).
\end{equation}

\begin{figure}
{\par\centering \resizebox*{0.8\columnwidth}{!}{\rotatebox{90}{
\includegraphics{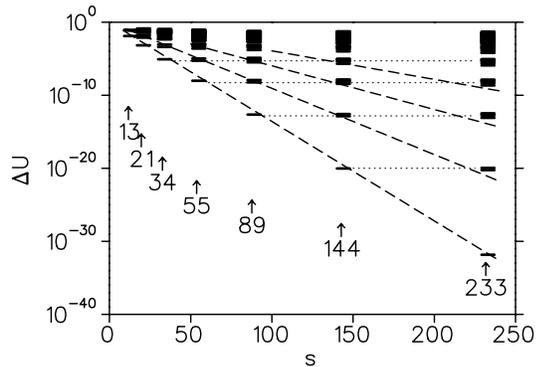}}} \par}
\vglue 0.2cm
\caption{\label{SzDep}Dependence of the band energies on the size $s$ of the chain for
\protect\( K=2\protect \) (\protect\( \lambda =0.7859)\protect \). Dashed
lines are given by empirical formula (\ref{bfit}), horizontal dotted lines allow to compare the band positions for different $s$.}
\end{figure}

According to Eqs.(\ref{bfit}),(\ref{bfitL}), the spacings between the bands and the ground state drops
exponentially with the length of the chain. 
In Fig.\ref{SzDep} we present the dependence of the 
band structure on the number of particles $s$ in the chain, for the  rational approximants $r/s$ of the golden mean \({\nu } \). 

The results presented in Figs. \ref{kdep},\ref{SzDep} show that the simple
Semi-empirical formula (\ref{bfit}),
shown by dashed lines, describes the positions of the bands in the interval of 30 orders of magnitude. It is interesting to note that bands are also ordered in some
horizontal levels (marked by dotted lines) which are practically independent of the size of the
chain. However, the band width and the number of states inside the band of the same horizontal level
grows with the chain size $s$. In the next section we will see how all these features
can be understood on the basis of the spatial properties of the chain structure.

\begin{figure}
{\par\centering \resizebox*{0.8\columnwidth}{!}{\rotatebox{90}{
\includegraphics{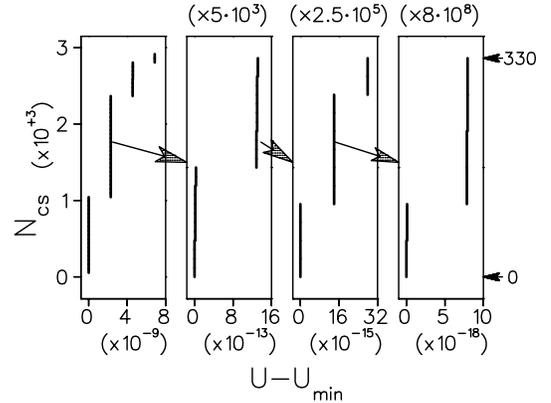}}} \par}
\vglue 0.2cm
\caption{\label{Fract}Fractal energy band structure for a chain with \protect\( K=4\protect \),
\protect\( s=89\protect \) and \protect\( r=55\protect \). Four hierarchical
levels with growing resolution are shown from left to right. The total magnification
factor in the horizontal direction is shown on the top. Here $U$ is the chain energy per particle and $U_{min}$ gives the energy of the most left band in each panel. The vertical scale $N_{cs}$ gives the integrated number of equilibrium configurations counted from the bottom of the most left band in each panel. The vertical magnification is changed in ten times from left to right.}
\end{figure}

Finally, in Fig.\ref{Fract} we show that the energy band structure is characterized by fractal properties.
Here the third excited band for the chain with $r/s =55/89$ and \( K=4 \) is shown with subsequently growing resolution (see magnification factors on the top of figure boxes). The hierarchical structure of the bands is evident. Such a structure becomes deeper and deeper with the increase of 
the chain length $s$. In the next section we show the origin of this structure
and develop a simple model to describe it.

\section{Spatial structure of equilibrium configurations.\label{SpatStruct}}

\subsection{Structure of the ground state.}

\begin{figure}
{\par\centering \resizebox*{0.8\columnwidth}{!}{\rotatebox{90}{
\includegraphics{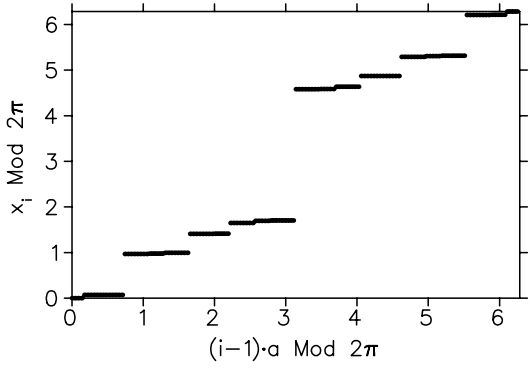}}} \par}
\vglue 0.2cm
\caption{\label{HullF}The particle position $x_i$ mod $2\pi$ versus the particle index (multiplied the average distance $a= 2\pi r/s$) mod $2\pi$. This hull function is shown for $r=89$,$s=144$ and $K=2$. }
\end{figure}

To analyze the  origin of the FK chain hierarchical structure let
us start with study of its \textit{ground state}. It is very instructive
to analyze regularities of particle positions inside the wells. In particular,
their positions modulo the period of the external field are given by the broadly
discussed hull function \cite{Au83a,Au83b,Au83c,Pe83}. Its typical example
is presented in Fig.\ref{HullF}.

\begin{figure}
{\par\centering \resizebox*{0.8\columnwidth}{!}{\rotatebox{90}{
\includegraphics{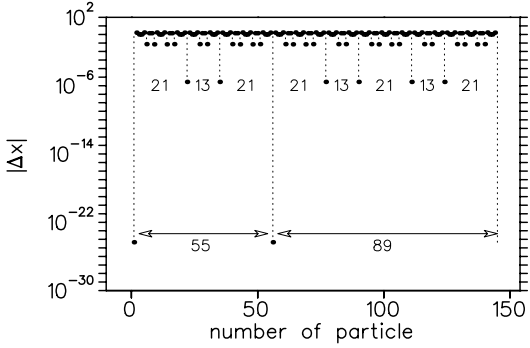}}} \par}
\vglue 0.2cm
\caption{\label{DevL}Absolute value of the particle deviations from the nearest potential well bottom 
 versus particle number along the chain. The chain
parameters are as in Fig.\ref{HullF}. }
\end{figure}

In this plot the bottoms of potential wells correspond to
 \( x_{i}\: {\rm Mod\, 2\pi }=0 \)
and \( 2\pi  \). It is easy to see, that a considerable amount of particles
is located very close to the bottoms. To render this observation even more significant,
the absolute values of deviations from the bottom versus the particle  number $i$ are plotted in logarithmic scale in Fig.\ref{DevL}.

We see, that some of particles are at the well bottoms  with extremely good accuracy!
Moreover, the values of small deviations are grouped into three well resolved
hierarchical levels. Separations along the chain for these particles are also
ordered in some regular way. The two particles closest to the bottom \( |\Delta x|\approx 4.7\cdot 10^{-25} \)
(!) are separated by the distances \( 55 \) and \( 89 \) (the chain is periodic).
Then 8 particles (including the previous two) whose deviation from the bottoms is \( |\Delta x|\leq 3\cdot 10^{-6} \),
 are separated by distances \( 13 \) and \( 21 \),
see Fig.\ref{DevL}. Finally, \( 34 \) particles whose deviation from bottoms is
\( |\Delta x|\leq 10^{-1} \) are separated by distances \( 3 \) and \( 5 \).
The greater is \( K \), the closer these particles are to the bottoms, yet their
separations along the chain remain the same. By taking a chain with longer length
(\( r/s=144/233,\, 233/377,\, 377/610,\ldots  \)) one can observe subsequent levels of the hierarchical structure.
 
The fact that some particles are very close to the bottom of the wells, is
very important. Indeed let us assume for a moment, that these particles are
\textit{exactly} at the well bottoms. This means, that tension forces acting
from both sides on any such particle, called hereafter as  {}``glue{}'' particle,
balance each other exactly. Now, let us cut the chain at glue particles into fragments,
or {}``bricks{}''. Then we can interchange any two fragments of the chain
without  changing  the chain potential energy. In general, the interchanged
bricks are different, and we get in this way a new configuration with the same
potential energy. So we may conclude, that we can get a combinatorially large
number of degenerate configurations in the ground state whose number grows  exponentially with the length of the chain.

In fact, our glue particles are lying very close to, but \textit{not
exactly}, the bottoms of wells. Actually they are slightly shifted from the bottoms, and therefore
the tensions \( f \) at the ends of different bricks are \textit{not} the same.
As a consequence, when we exchange two different bricks, each  brick's end will be slightly distorted. The distortion is
proportional to the difference in boundary tensions \( \Delta f \) of the nearby 
bricks. This leads to a local change of the chain
energy

\begin{equation}
\label{DUloc}
\Delta U\sim \Delta U_{b}+\Delta U_{g},
\end{equation}

where \( \Delta U_{b}\sim (\Delta f)^{2}/2 \) is due to the distortion of nearby 
bricks, and \( \Delta U_{g}\approx K(\Delta x)^{2}/2=(\Delta f)^{2}/2K \) is the change of potential energy due to the shift of the glue particle between the
bricks. 
We note that, since glue particle deviations are exponentially small and hierarchically
ordered, then the corresponding tension differences are also exponentially small and ordered. Therefore the energy change caused by bricks permutation depends on the level of the hierarchy inside which the permutation is done.
The lowest level of the hierarchy is built by bricks of two types, which 
consist of two and four particles respectively. 
For brevity let us denote them as \( 2 \) and \( 4 \). Then a chain which consist of \( 8 \) particles can be denoted  as \( g2g4 \) (the letter $g$ stands for a glue particle). The tension difference  at this level of the hierarchy is \( \Delta f=K\, \Delta x\sim 10^{-1} \).

The next level of hierarchy has bricks \( 12=(4g2g4) \) and \( 20=(4g2g4g2g4) \).
The brackets are introduced for convenience to denote the form of the brick. The tension difference at this level is
much smaller: \( \Delta f=K\, \Delta x\sim 10^{-5} \). Finally, the
third level of the hierarchy is composed in the similar way: \( 54=(20g12g20) \) and
\( 88=(20g12g20g12g20) \), with the corresponding tension difference \( \Delta f=K\, \Delta x\sim 10^{-24} \). With increasing the particle number the above described process proceeds in a similar way. A simple estimate for the tension difference valid at any hierarchical level, and
for any \( K \),  can be written as:
\( \Delta f\sim K\exp (-\lambda s_{min}) \), where \( s_{min} \) is the number
of particles in the smallest brick at the given level of hierarchy and $\lambda$   is the phonon gap which depends implicitly on $K$.
Notice that a brick with the addition of the glue particle, forms an elementary cell the size of which is given by the Fibonacci numbers. For example (g2)=3,(g4)=5, (g2g4)=8 etc.

The composition rules for the brick construction at any hierarchical level can be summarized in following way. Suppose that a
 given level of hierarchy is composed by two bricks \( A \) and \( B \),
with the length of \( A \) smaller than $B$. Then the bricks \( A',\, B' \) of
the next level can be built as

\begin{equation}
\label{CmpRul}
A'=BgAgB,\quad B'=BgAgBgAgBg
\end{equation}

In principle, the composition rules just described allow to build the ground state
for a chain of any length. It is also clear that for long enough chains one does not need  to search the global minimum of the potential energy. Instead, it is sufficient
to minimize the energy of bricks up to some hierarchical level:  any further
optimization goes beyond any reasonable precision. This however also means
that within the same  precision the  ground state configuration described by Aubry is indistinguishable from exponentially many  \textit{disordered}
excited configurations.

\subsection{Structure of the excited configurations.}

\begin{figure}
{\par\centering \resizebox*{0.8\columnwidth}{!}{\rotatebox{90}{
\includegraphics{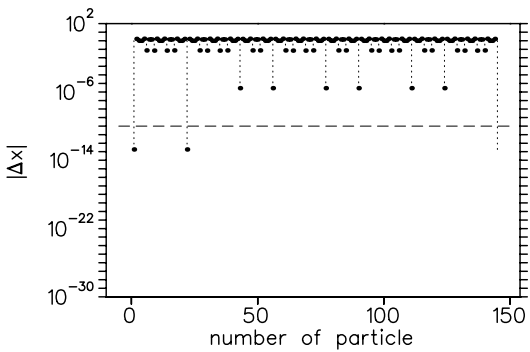}}} \par}
\vglue 0.2cm
\caption{\label{DevBex}Deviations of the particles from potential well bottoms for an excited configuration taken from the first excited energy band (see Fig.\ref{SzDep} for $s =144$). The separation of this band from the ground state of Fig.\ref{HullF} is
\protect\( \Delta U_{1}\sim 10^{-20}\protect \), which implies that the deviations of glue particles from the well bottoms are
 \protect\( \Delta x\leq 10^{-10}\protect \)(dashed line). }
\end{figure}

\begin{figure}
{\par\centering \resizebox*{0.8\columnwidth}{!}
{\rotatebox{90}
{\includegraphics{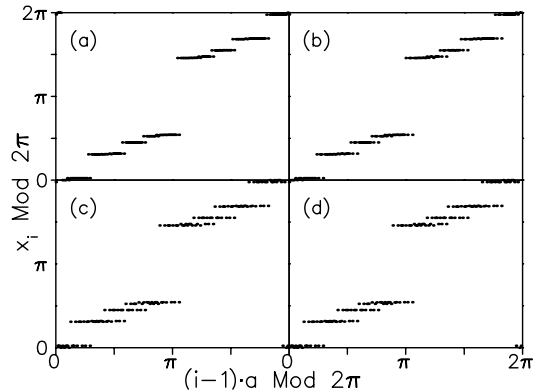}}} \par}
\vglue 0.2cm
\caption{\label{Hullex} The hull function for a typical equilibrium 
configuration in the $k$-th excited band for $K=2$ and $r/s =89/144$: 
a)$k=1$, b) $k=2$, c) $k=3$, d) $k=4$.
The energy band structure is shown in Fig. \ref{SzDep}. 
Compare with the hull function of the ground state shown in 
Fig. \ref{HullF}.
}
\end{figure}

The picture of the ground state described above allows also to understand the structure of excited configurations. However, in this case the structure can be
a bit less self-evident. To illustrate this, in Fig.\ref{DevBex}
we plot particles deviations from well bottoms for a configuration from the first  excited band in the chain shown in Fig.\ref{DevL}.
The hull function for a typical configuration in this band is shown in Fig.\ref{Hullex}(a). The hull function for a typical configuration in the second, third and fourth excited bands (see the band structure in Fig.\ref{SzDep} with $s=144$) is shown in Fig.\ref{Hullex}.
Contrary to the monotonic hull function of the ground state, here the hull function becomes not monotonic and one can see the overlap between horizontal plateaus. 

From Fig.\ref{DevBex} we see that for the first two levels of hierarchy, the deviations of glue particles from the well bottom are practically the same as in the ground state (see Fig. \ref{DevL}). However at the third hierarchical level the deviations
of two glue particles (below the dashed line)  become considerably larger
than the  corresponding ones in the ground state (see Fig.\ref{DevL}).

In order to give an unambiguous definition of bricks  
let us remind, that we want to split the chain into bricks, which permutations
keeps the chain configuration inside the same band. According to Eq.(\ref{DUloc}),
the energy change due to a permutation, produced by the  tension differences between permuted bricks, can be estimated as \( \delta U\sim (\Delta x)^{2} \).
Therefore the deviations of glue particles from the bottom between
the  bricks is restricted by the condition \( \Delta x\leq (\Delta U_{k})^{1/2} \),
where \( \Delta U_{k}\) is the band energy counted from the ground
state. Taking this condition into account, we can write for the configuration shown  in Fig.\ref{DevBex} its decomposition into bricks as \( g20g122=g20g(20g12g20g12g20g12g20) \),
where the expansion of the configuration is shown  up to bricks of the second level, \( 12 \)
and \( 20 \). As above, by brackets we mark the chain fragments which in  permutations
should be considered as a single brick, since their destruction results in the
energy change exceeding the band width. 

Let us now discuss the properties of the bricks expansion on the example of a periodical  chain with $r/s= 89/144$ and $K=2$ (see Fig.\ref{SzDep}). The first excited band $k=1$ has the excitation energy \( \Delta U_1\approx 9.16\cdot 10^{-21} \) and is composed from one configuration \( g20g(20g12g20g12g20g12g20) \)(here we do not count the configurations with a shift along the chain and reflection). It is interesting to note that this configuration has a long commensurate fragment (123/144).

The second excited band $k=2$ has energy \( \Delta U_2\lesssim 10^{-12} \).

It is composed by three configurations: \( g12g20g20g20g12g20g20g12 \), \( 12g20g20g20g20g12g20g12 \)
and \( g20g20g20g20g20g12g12g12 \). With the configuration from the first band they give all possible different combinations of three bricks \( 12 \) and five bricks
\( 20 \) which are used in the composition of the ground state. 

The third band $k=3$ has the excitation energy \( \Delta U_3\lesssim 10^{-8} \). This band has too many configurations to be listed here. Let us however mention a new phenomenon which appears in this band namely a
 brick {}``chemical{}'' reaction with dissociation of \textit{larger}
elementary bricks of the second hierarchical level:

\begin{equation}
\label{dis20}
20+20\, \rightarrow \, 12+28,\quad 20+28\, \rightarrow \, 12+36 .
\end{equation}

Note, that a {}``free radical{}'' \( 8 \) coming from dissociation
\( 20\, \rightarrow \, 12+8 \) is easily captured by other long bricks, so that there is a considerable contribution of long commensurate structures.
Near the bottom of the band a typical configuration is \( g20g12g29g12g20g12g12g20 \),
While at the top one has \( g53g12g12g12g12g12g12 \)!

The fourth band $k=4$ has energy \( \Delta U_4\lesssim 10^{-5} \). Here we see a dissociation
of the bricks from the second hierarchical level:

\begin{equation}
\label{dis12}
12+20\, \rightarrow \, 4+28,\quad 12+28\, \rightarrow \, 4+36,\ldots ,
\end{equation}

and appearance of elementary bricks $4$ from  the first hierarchical level.
Here are some examples of configurations in this band with bricks $4$:
 \( g20g20g28g20g12g12g4g20 \),
\( g12g12g12g28g12g20g4g36,\ldots  \). 

Further steps in the whole picture are straightforward. Now we outline
a simple theory which turns our qualitative observations into quantitative predictions for the band energy spectrum.

\subsection{An analytical approach.}

\label{theory}

In fact, the construction of bricks is based on the existence of an intrinsic small parameter which
allows to develop a simple rapidly converging perturbation theory. Here we
 outline its main elements. Let us consider the FK chain with $s$ particles and fixed ends at \( x_{0}=0 \) and \( x_{s}=2\pi r \). Then the largest brick contains $n= s-1$ particles. 
If the  glue particles ($i=0, i=s$) are slightly shifted from the well bottoms \( x_{a,b}\ll <1 \),
then the brick energy can be written as

\begin{equation}
\label{UBexp}
\begin{array}{rcl}
U^{(n)}(x_{a},x_{b}) & = & U^{(n)}_{0}-f^{(n)}_{a}x_{a}+f^{(n)}_{b}x_{b}+\\
 &  & +R_{a}^{(n)}\frac{x_{a}^{2}}{2}+R_{b}^{(n)}\frac{x^{2}_{b}}{2}-T^{(n)}x_{a}x_{b},
\end{array}
\end{equation}

where \( U^{(n)}_{0}=U^{(n)}(0,0) \) is the unperturbed energy, \( f^{(n)}_{a,b} \) and \( R^{(n)}_{a,b} \)
are tensions and rigidities at the left/right ends of the brick, and \( T^{(n)} \)
is the static {}``transmission{}'' factor along the brick with \( n \)particles. If the brick
is symmetric then \( f^{(n)}_{a}=f^{(n)}_{b}\equiv f^{(n)} \) and \( R^{(n)}_{a}=R^{(n)}_{b}\equiv R^{(n)} \). 
The key point of the theory is that in
the presence of a nonzero phonon gap \( \lambda  \) the transmission factor \( T^{(n)} \) is exponentially small:
\( T^{(n)}\sim \exp (-\lambda n) \). Therefore it can be very efficiently used as an
expansion parameter in the calculations of the energy band spectrum.

Suppose that at some  hierarchical level we have two elementary bricks \( A \)
and \( B \), with lengths \( n_{A}<n_{B} \). According to our rule of brick
composition (\ref{CmpRul}), we can calculate the energy  of the brick \( A'=BgAgB \) as:

\begin{equation}
\label{UAp}
U^{(A')}(x_{a},x_{b})=\min _{x_{1}x_{2}}[U^{B}(x_{a},x_{1})+U^{A}(x_{1},x_{2})+U^{B}(x_{2},x_{b})].
\end{equation}

Then, rewriting (\ref{UAp}) in the form (\ref{UBexp}) we obtain the transformation
rules for brick parameters \( R, \) \( f \) and \( T \). In the leading order approximation in the small parameter \( T \) these rules have the form: 

\begin{equation}
\label{trA}
\begin{array}{lcl}
R^{A'} & = & R^{A}\\
f^{A'} & = & f^{B}-\frac{T^{B}(f^{B}-f^{A})}{R^{A}+R^{B}+K}+\cdots \\
T^{A'} & = & \frac{T^{A}(T^{B})^{2}}{(R^{A}+R^{B}+K)^{2}}.
\end{array}
\end{equation}

In the same way for \( B' \) we obtain

\begin{equation}
\label{trB}
\begin{array}{lcl}
R^{B'} & = & R^{B}\\
f^{B'} & = & f^{A'}+(\Delta f)^{'}\\
T^{B'} & = & \frac{T^{B}(T^{A}T^{B})^{2}}{(R^{A}+R^{B}+K)^{4}}
\end{array}
\end{equation}

where the tension difference \( (\Delta f)'=f^{B'}-f^{A'} \) between new bricks
\( A' \) and \( B' \) can be expressed through the brick tension difference (\( \Delta f)=f^{B}-f^{A} \)
as: 

\begin{equation}
\label{fDif}
(\Delta f)'=-\frac{T^{A}(T^{B})^{2}}{(R^{A}+R^{B}+K)^{2}}(\Delta f).
\end{equation}

To apply these transformation rules one needs to know the bricks parameters at the
lowest hierarchical level, e.g. for bricks \( 2 \) and \( 4 \). In this case  the number
of particles is small and the expansion (\ref{UBexp}) can be performed  analytically.
For the case \( K=2 \) considered above, we get \( T^{(2)}=0.24, \)
\( R^{(2)}=0.454 \), \( T^{(4)}=0.0533, \) \( R^{(4)}=0.32 \), and \( (\Delta f)^{(2,4)}=0.21 \).
By applying the transformation rules to these data we obtain for the bricks of the next hierarchical level
 \( 12 \) and \( 20 \): \( T^{(12)}=8.86\cdot 10^{-5} \), \( T^{(20)}=1.796\cdot 10^{-7} \),
\( (\Delta f)^{(12,20)}=-6.93\cdot 10^{-6} \). The exact numerical simulation gives
\( T^{(12)}=9.82\cdot 10^{-5} \), \( T^{(20)}=1.799\cdot 10^{-7} \), \( (\Delta f)^{(12,20)}=-7.16\cdot 10^{-6} \).
Starting with exact values for bricks \( 12 \) and \( 20 \) the transformation
rules give for bricks \( 54 \) and \( 89 \) results which are correct within
 four digits accuracy!

Therefore this simple approach can quantitatively explain the splitting
of the whole spectra into bands. Surely, the leading terms  in the
small parameter \( T \), as well as the expansion  (\ref{UBexp}), can be insufficient
to reproduce with high accuracy the deep levels of hierarchical band structure. To this end one should take into account higher order terms.

\begin{figure}
{\par\centering \resizebox*{0.8\columnwidth}{!}{\rotatebox{90}{
\includegraphics{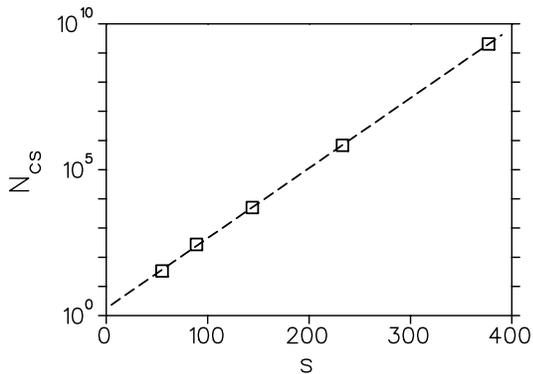}}} \par}
\vglue 0.2cm
\caption{\label{Ncsexp} The number of equilibrium configurations $N_{cs}$ with excitation energy from the ground state $\Delta U \leq 2\cdot 10^{-8}$ as a function of the number of particles $s$ in the chain for $K=2$. The dashed line shows the fitted exponential dependence $N_{cs} = 1.78 \exp(0.0554 s)$.}
\end{figure}

The results presented in this section show that the number of equilibrium configurations grows very quickly with the length of the chain and with the chaos parameter $K$. These configurations form bands which are placed exponentially close to the ground state. As a result, even in a fixed very small vicinity of the ground state, the number of configurations grows exponentially with the chain length. This fact is illustrated in Fig. \ref{Ncsexp}.

\section{Discussion and conclusions.}

In this paper we studied the properties of equilibrium static configurations in the Frenkel-Kontorova chain in the regime of pinned phase characterized by phonon gap. This FK model is rather general and finds applications not only for commensurate -incommensurate transition for atoms placed on a periodic substrate but also in many other fields of physics. 
In addition, near the equilibrium, also 
 the cases with long range interactions between atoms  can be effectively reduced to the FK model 
with only nearest neighbors interaction. We have shown that energies of equilibrium configurations form a hierarchical band structure so that exponentially many configurations become exponentially close to the unique ground state. In this respect the FK model has certain similarities with classical spin glass models which also are characterized by existence of exponentially many quasi-degenerate states \cite{Parisi}. At the same time in the FK model the disorder is absent and the quasi-degenerate configurations form a fractal sequence of energy bands which in a sense can be considered a dynamical spin glass. On the basis of extended numerical and analytical investigations we determined the low energy excitation inside the quasi-degenerate bands which have a form of bricks from which the whole chain can be composed. On the basis of these results we have shown that while the ground state is characterized by regular structure, the low energy excited configurations are disordered due to elementary brick  
displacements. This means that exponentially close to the ground state there are disordered configurations which may have rather different physical properties compared to the ground state. For example this disorder should significantly affects the properties of phonon excitations in the chain. 
The exponential quasi-degeneracy of low energy configurations should be also important in the case of quantum FK chain when quantum particles can tunnel from one configuration to another. These two aspects are related with new interesting physical effects of low energy excitations in many-body systems and require further investigations \cite{Casa00}.  

This work was supported in part by the EC RTN network contract HPRN--CT-2000-0156. One of us (O.V.Z.) thanks
Cariplo fundation, INFN and RFBR grant No. 01-02-17621 for financial support.
Support from the PA INFN ``Quantum transport and classical chaos'' is gratefully acknowledged.

\end{document}